\documentclass[letter,bibyear]{aa} 

\usepackage{epsfig}
\usepackage{amsmath}
\usepackage{subfigure}
\usepackage{natbib}
\usepackage{multirow}
\usepackage{color}
\usepackage{longtable}
\usepackage{graphicx}
\usepackage{epstopdf}
\usepackage{booktabs}
\usepackage{txfonts}

\newcommand{\jmst}{J.~Mol.~Struct.}

\newcommand{\kms}{km s$^{-1}$}
\newcommand{\diez}{10$^{10}$\,cm$^{-2}$}
\newcommand{\once}{10$^{11}$\,cm$^{-2}$}
\newcommand{\doce}{10$^{12}$\,cm$^{-2}$}
\newcommand{\trece}{10$^{13}$\,cm$^{-2}$}
\begin{document}

\title{Discovery of  CH$_2$CCHC$_4$H and a rigorous detection of CH$_2$CCHC$_3$N in TMC-1 with the QUIJOTE$^1$ line survey\thanks{Based on 
observations carried out
with the Yebes 40m telescope (projects 19A003,
20A014, 20D023, 21A011, and 21D005). The 40m
radiotelescope at Yebes Observatory is operated by the Spanish Geographic Institute
(IGN, Ministerio de Transportes, Movilidad y Agenda Urbana).}}

\author{
R.~Fuentetaja\inst{1},
C.~Cabezas\inst{1},
M.~Ag\'undez\inst{1},
B.~Tercero\inst{2,3},
N.~Marcelino\inst{3},
J.~R.~Pardo\inst{1},
P.~de~Vicente\inst{2}
J.~Cernicharo\inst{1},
}

\institute{Dpt. de Astrof\'isica Molecular, Instituto de F\'isica Fundamental (IFF-CSIC),
C/ Serrano 121, 28006 Madrid, Spain. \email r.fuentetaja@csic.es, jose.cernicharo@csic.es
\and Centro de Desarrollos Tecnol\'ogicos, Observatorio de Yebes (IGN), 19141 Yebes, Guadalajara, Spain.
\and Observatorio Astron\'omico Nacional (OAN, IGN), Madrid, Spain.
}

\date{Received; accepted}

\abstract{Using the QUIJOTE$^1$ line survey in the 32.0-50.4 GHz range, we report the discovery 
of the molecule CH$_2$CCHC$_4$H towards the prestellar cold core TMC-1 in the Taurus region. 
We also present a rigorous detection of CH$_2$CCHC$_3$N,  along with its detailed analysis. We identified a total of twenty rotational transitions for each one of these molecules. The 
rotational quantum numbers range from $J_u$=17 up to 24 and $K_a\le$3. 
The column density for CH$_2$CCHC$_4$H 
is $N$=(2.2$\pm$0.2)$\times$10$^{12}$ cm$^{-2}$, while for CH$_2$CCHC$_3$N, we derived 
$N$=(1.2$\pm$0.15)$\times$10$^{11}$ cm$^{-2}$. The rotational temperature is 9.0$\pm$0.5 K for
both species. The abundance ratio between
CH$_2$CCHC$_4$H and CH$_2$CCHC$_3$N is 18$\pm$4. We also compared the column densities of these
species with those of their isomers CH$_3$C$_6$H and CH$_3$C$_5$N, derived from their $J$=20-19 up to
J=30-29 rotational transitions observed with the QUIJOTE line survey. The observed abundances for all
these species are reasonably well explained by state-of-the-art chemical models of TMC-1.
The observed astronomical frequencies were merged with laboratory frequencies from 
the literature to derive improved spectroscopic parameters.
}


\titlerunning{CH$_2$CCHC$_4$H in TMC-1}
\authorrunning{Fuentetaja et al.}

\maketitle

\section{Introduction}
The recent results obtained towards TMC-1 with the Yebes 40m radio telescope through the 
QUIJOTE$^1$ line survey \citep{Cernicharo2021a} and those of the
Green Bank 100m telescope with the GOTHAM survey \citep{McGuire2018}
show that this source presents a paramount challenge to 
our understanding of the chemical processes in cold prestellar cores.
Several cyclic molecules have been detected such as cyclopentadiene and indene \citep{Cernicharo2021b}, as well as
ortho-benzyne \citep{Cernicharo2021a}, the cyano
derivatives of cyclopentadiene, benzene, and naphthalene \citep{McGuire2018,McCarthy2021,
Lee2021,McGuire2021}, and the ethynyl derivatives of cyclopentadiene \citep{Cernicharo2021c}. Moreover,
several extremely abundant hydrocarbons such as 
propargyl \citep{Agundez2021}, vinyl acetylene \citep{Cernicharo2021d}, and allenyl acetylene \citep{Cernicharo2021e}
have also been discovered towards TMC-1. 
Most of these species had not been predicted to exhibit a significant abundance in chemical models
prior to their detections. These results indicate
that key chemical processes involving hydrocarbons (radical, neutral, 
ion-neutral, and perhaps dust grain surface reactions) have been neglected in previous
chemical networks. Thus, a new chemistry is emerging based on these TMC-1 observations.

The study of ethynyl and cyanide  derivatives of hydrocarbons is of paramount importance, as they can allow us to follow the different chemical paths to produce them from a common
precursor through reactions involving the CCH and CN radicals \citep{Cernicharo2021f}. In this regard, the discovery in TMC-1 of various isomers with formulae C$_5$H$_4$ and C$_4$H$_3$N allows us to also constrain the chemical schemes behind their formation \citep{Cernicharo2021e,Marcelino2021}. These schemes involve reactions of CCH and CN with the hydrocarbons CH$_3$CCH and CH$_2$CCH$_2$,  allowing for an indirect estimation of the abundance of the non-polar but abundant hydrocarbon allene to be made. The same chemical schemes leading to C$_5$H$_4$ and C$_4$H$_3$N isomers can be extended to longer species by substituting the radicals CCH and CN by their longer equivalents C$_4$H and C$_3$N, which would result in different isomers with formula C$_7$H$_4$ and C$_6$H$_3$N.

In this letter, we report the discovery of CH$_2$CCHC$_4$H (allenyl diacetylene) using
a line survey performed with the Yebes 40m telescope (QUIJOTE\footnote{\textbf{Q}-band \textbf{U}ltrasensitive \textbf{I}nspection \textbf{J}ourney 
to the \textbf{O}bscure \textbf{T}MC-1 \textbf{E}nvironment}; see \citealt{Cernicharo2021a}). This species is an isomer of CH$_3$C$_6$H, both with the formula C$_7$H$_4$. We also present a detailed study and a rigorous and robust detection of CH$_2$CCHC$_3$N, an isomer of CH$_3$C$_5$N (both with formula C$_6$H$_3$N) which was recently claimed through stacking techniques by \citet{Shingledecker2021}. Each of these large species has small rotational constants. These, together with an unbiased line surveys such as QUIJOTE$^{1}$, allow studying many rotational transitions and, hence, to derive homogeneous and realistic values of their column densities. We investigate the chemistry of these species with the aid of updated gas-phase chemical models.

\begin{figure*}
\centering
\includegraphics[width=1.0\textwidth]{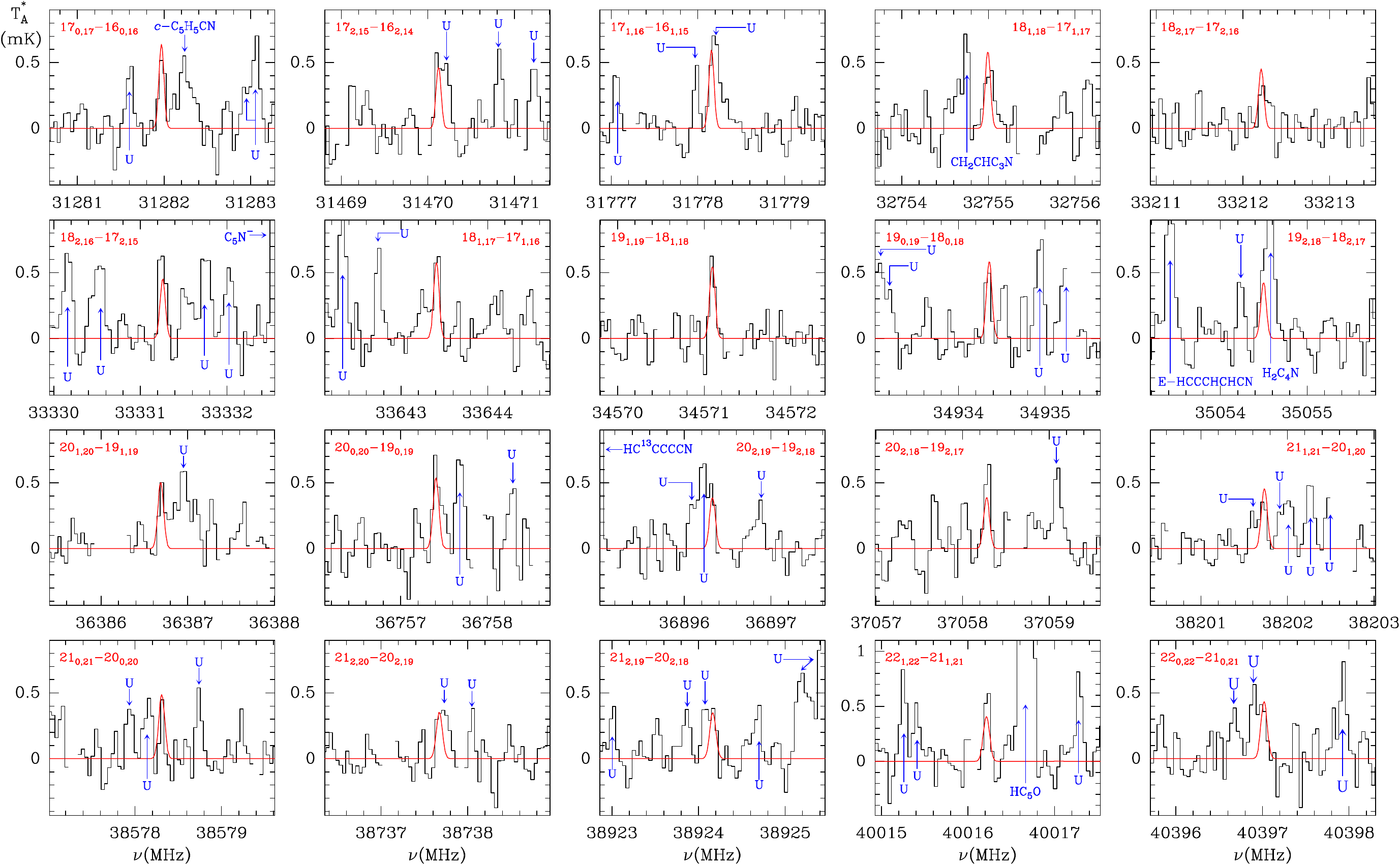}
\caption{Observed transitions of CH$_2$CCHC$_4$H in TMC-1.
The abscissa corresponds to the rest frequency of the lines. Frequencies and intensities for the observed lines
are given in Table \ref{obs_line_parameters}.
The ordinate is the antenna temperature, corrected for atmospheric and telescope losses, in milli Kelvin.
The quantum numbers of each transition are indicated
in the corresponding panel. Red line shows the computed synthetic spectrum for this species for $T_{\mathrm{rot}}$ = 9 K and
a column density of 2.2$\times$\doce. Blanked channels correspond to negative features 
produced when folding the frequency-switched data.
}
\label{fig_ch2cchc4h}
\end{figure*}

\section{Observations}
\label{observations}
New receivers built within the Nanocosmos project\footnote{\texttt{https://nanocosmos.iff.csic.es/}}
and installed at the Yebes 40m radiotelescope were used
for the observations of TMC-1
($\alpha_{J2000}=4^{\rm h} 41^{\rm  m} 41.9^{\rm s}$ and $\delta_{J2000}=
+25^\circ 41' 27.0''$). A detailed description of the system is 
given by \citet{Tercero2021}.
The receiver consists of two cold high electron mobility transistor amplifiers covering the
31.0-50.3 GHz band with horizontal and vertical             
polarizations. Receiver temperatures in the runs achieved during 2020 vary from 22 K at 32 GHz
to 42 K at 50 GHz. Some power adaptation in the down-conversion chains effectively reduced
the receiver temperatures during 2021 to 16\,K at 32 GHz and 30\,K at 50 GHz.
The backends are $2\times8\times2.5$ GHz fast Fourier transform spectrometers
with a spectral resolution of 38.15 kHz,
providing full coverage of the Q-band in both polarizations.

A detailed description of the QUIJOTE line survey is
provided in \citet{Cernicharo2021a}. In brief, the present data
come from several observing runs between December 2019 and January 
2022. 
All observations were performed in the frequency-switching mode with frequency throws of 8 and 10 MHz. 
The total on-source observing time for data taken with
frequency throws of 8 MHz and 10 MHz is 173.1 and 253.6 hours, 
respectively. Hence, the total observing time on source by
December 2021 is 426.7 hours.
The measured sensitivity varies between 0.12 mK at 32 GHz and
0.25 mK at 49.5 GHz.
For each frequency throw, different local oscillator
frequencies were used in order to remove possible side 
band effects in the down conversion chain.

The main beam efficiency varies from 0.6 at
32 GHz to 0.43 at 50 GHz \citep{Tercero2021}. 
Pointing corrections were derived from nearby quasars and SiO masers
and the errors were always within 2-3$''$. The telescope 
beam size is 56$''$ and 31$''$ at 31 and 50 GHz, respectively.
The intensity scale used in this work, antenna temperature
($T_A^*$), was calibrated using two absorbers at different temperatures and the
atmospheric transmission model ATM \citep{Cernicharo1985, Pardo2001}.
The adopted calibration uncertainties are at the level of  10\% and
all the data were analysed using the GILDAS package\footnote{\texttt{http://www.iram.fr/IRAMFR/GILDAS}}.

\begin{figure*}
\centering
\includegraphics[width=1.0\textwidth]{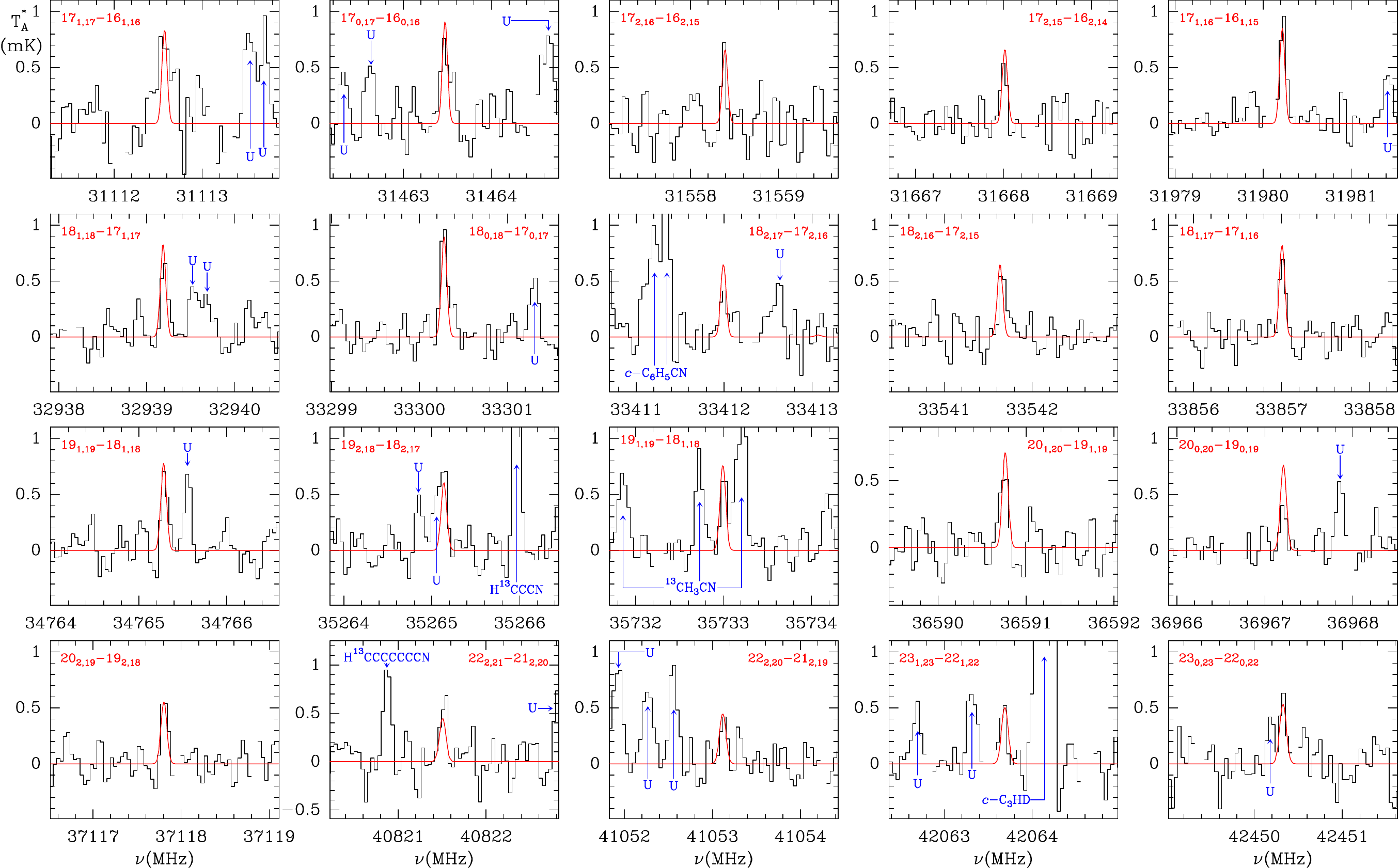}
\caption{Selected transitions of CH$_2$CCHC$_3$N in TMC-1.
The abscissa corresponds to the rest frequency of the lines. Frequencies and intensities for all observed lines
are given in Table \ref{obs_line_parameters}.
The ordinate is the antenna temperature, corrected for atmospheric and telescope losses, in milli Kelvin.
The quantum numbers of each transition are indicated
in the corresponding panel. The red line shows the computed synthetic spectrum for this species for $T_{\mathrm{rot}}$ = 9 K and
a column density of 1.2$\times$\once. Blanked channels correspond to negative features 
produced when folding the frequency-switched data.
}
\label{fig_ch2cchc3n}
\end{figure*}

\section{Results}
\label{results}
The line identification in this work has been achieved using the 
MADEX \citep{Cernicharo2012}, CDMS \citep{Muller2005}, and JPL \citep{Pickett1998} catalogues. 
By April 2021, the MADEX has featured 6433 spectral entries corresponding to the ground and vibrationally excited states, together
with the corresponding isotopologues of 1733 molecules.  

\subsection{Discovery of CH$_2$CCHC$_4$H}
The search for new species
in our survey is based on intensity predictions for the molecules for which their rotational
spectroscopy is available. The sensitivity of QUIJOTE is a factor $\sim$50 larger 
than that of previous TMC-1 Q-band line surveys \citep{Kaifu2004}, which has
permitted us to address the search of molecules with a very low dipole moment \citep{Cernicharo2021a}.
Taking into account the large abundances derived from QUIJOTE's data for
hydrocarbons such as CH$_2$CHCCH \citep{Cernicharo2021d}, CH$_2$CCH \citep{Agundez2021,Agundez2022}, 
CH$_2$CCHCCH \citep{Cernicharo2021e}, H$_2$C$_5$ \citep{Cabezas2021}, c-C$_6$H$_4$ \citep{Cernicharo2021a}, 
c-C$_5$H$_6$, and c-C$_9$H$_8$ \citep{Cernicharo2021d}, other
similar species could be present in the survey, such as allenyl diacetylene (CH$_2$CCHC$_4$H).
Rotational spectroscopy for this species has been measured in the laboratory by
\citet{McCarthy2020}. We fitted these lines and implemented the rotational
constants into MADEX \citep{Cernicharo2012}. The derived constants, given in Table 
\ref{new_rotational_ch2cchc4h}, are identical to those of \citet{McCarthy2020}. We predicted the intensities for
all lines above 0.5 mK, adopting an initial column density of 5$\times$\doce and a rotational temperature of 10\,K. Twenty lines of this
species with intensities between 0.4 and 0.7 mK have been found with a signal-to-noise ratio (S/N)
between 3 and 6. These are shown on Fig. \ref{fig_ch2cchc4h} and
the derived line parameters are given in Table \ref{obs_line_parameters}. Several lines are partially blended
with other features, most of them arising from unknown species. Nevertheless, many lines appear as
single features and provide a solid detection of this molecule. 

Using the observed frequencies and
those measured in the laboratory, we derived a new set of rotational and distortion constants.
They are provided in section \ref{new_constants} and Table \ref{new_rotational_ch2cchc4h}. The difference between observed and calculated values are given in Table \ref{obs_frequencies_ch2cchc4h}. The new constants are practically
identical to those of \citet{McCarthy2020}, but have lower uncertainties, which allow
for more precise frequency predictions at higher frequencies.

To derive the column density and rotational temperature we used a model line fitting procedure 
(see e.g. \citealt{Cernicharo2021d}), which
leads to N(CH$_2$CCHC$_4$H)=(2.2$\pm$0.2)$\times$10$^{12}$ cm$^{-2}$ and $T_{\mathrm{rot}}$=9.0$\pm$0.5 K.
A rotational temperature lower than 8.5\,K results on an underestimation of the observed 
intensity for transitions with $K_a$=2. We assumed a source of uniform brightness with
a diameter of 80$''$ \citep{Fosse2001,Cernicharo2022}. The predicted synthetic lines for these parameters
are shown on Fig. \ref{fig_ch2cchc4h} (red lines). Adopting a H$_2$ column density for TMC-1 of 10$^{22}$
cm$^{-2}$ \citep{Cernicharo1987}, the abundance of CH$_2$CCHC$_4$H is 
(2.2$\pm$0.2)$\times$10$^{-10}$. Using a CH$_2$CCHCCH
column density of $N$=1.2$\times$\trece\,\citep{Cernicharo2021e},  we 
derive a CH$_2$CCHCCH/CH$_2$CCHC$_4$H abundance ratio of $\sim$5.5, similar
to the  CCH/C$_4$H radicals abundance ratio of $\sim$4 derived by \citet{Cernicharo2022}.

\subsection{A robust detection of CH$_2$CCHC$_3$N}
CH$_2$CCHC$_3$N (cyanoacetyleneallene or 4,5-hexadien-2-ynenitrile) was recently reported as having been detected towards TMC-1 by \citet{Shingledecker2021}. No individual lines were reported
but the stacking of all lines within their line survey. In order to search for the
lines of this species with QUIJOTE, we fit the rotational lines of this
species kindly provided to us by M. McCarthy. Using
these constants, we predicted the frequencies for all the lines of this molecule with $K_a\le$3 within
the Q-band. Twenty eight lines were detected and their line parameters are given in Table \ref{obs_line_parameters}.
Twenty of these lines are shown on Fig. \ref{fig_ch2cchc3n}. Although some lines are blended, the detection is now rigorous, as many individual lines are clearly seen in our data. Improved rotational and distortion constants
are provided in section \ref{new_constants} and Table \ref{new_rotational_ch2cchc3n}.

We assumed the same source parameters than for CH$_2$CCHC$_4$H. Through a model fitting procedure, we derive a rotational temperature of
9.0$\pm$0.5 K and a column density of (1.20$\pm$0.15)$\times$\once. Hence, the abundance of CH$_2$CCHC$_3$N
is (1.20$\pm$0.15)$\times$10$^{-11}$. Our column density agrees with the total value derived
from the four velocity and source components of model B of \citet{Shingledecker2021} (1.6$\times$\once, see their Table A.1).
The results concerning their models A or C can be fully discarded, as they provide unreasonable rotational temperatures (model A) or extremely large column densities (model C). 
We consider that our determination based on the observation of many individual lines is more accurate. Hence, the CH$_2$CCHC$_4$H/CH$_2$CCHC$_3$N abundance ratio is 18$\pm$4. 

It is interesting to compare the column densities of the different ethynyl and cyanide derivatives
of species containing the allenyl group CH$_2$CCH. To this purpose, we derived the column
density of C$_4$H and C$_3$N from our data, and we found N(C$_4$H)=1.1$\times$10$^{14}$cm$^{-2}$ and N(C$_3$N)=1.4$\times$10$^{13}$cm$^{-2}$ \citep{Cernicharo2022}. These values result in an abundance ratio C$_4$H/C$_3$N of 8, which is somewhat lower than the CH$_2$CCHC$_4$H/CH$_2$CCHC$_3$N ratio of 18$\pm$4 derived in this work.

\section{Discussion} 
\label{discussion}

\begin{figure}
\centering
\includegraphics[width=\columnwidth]{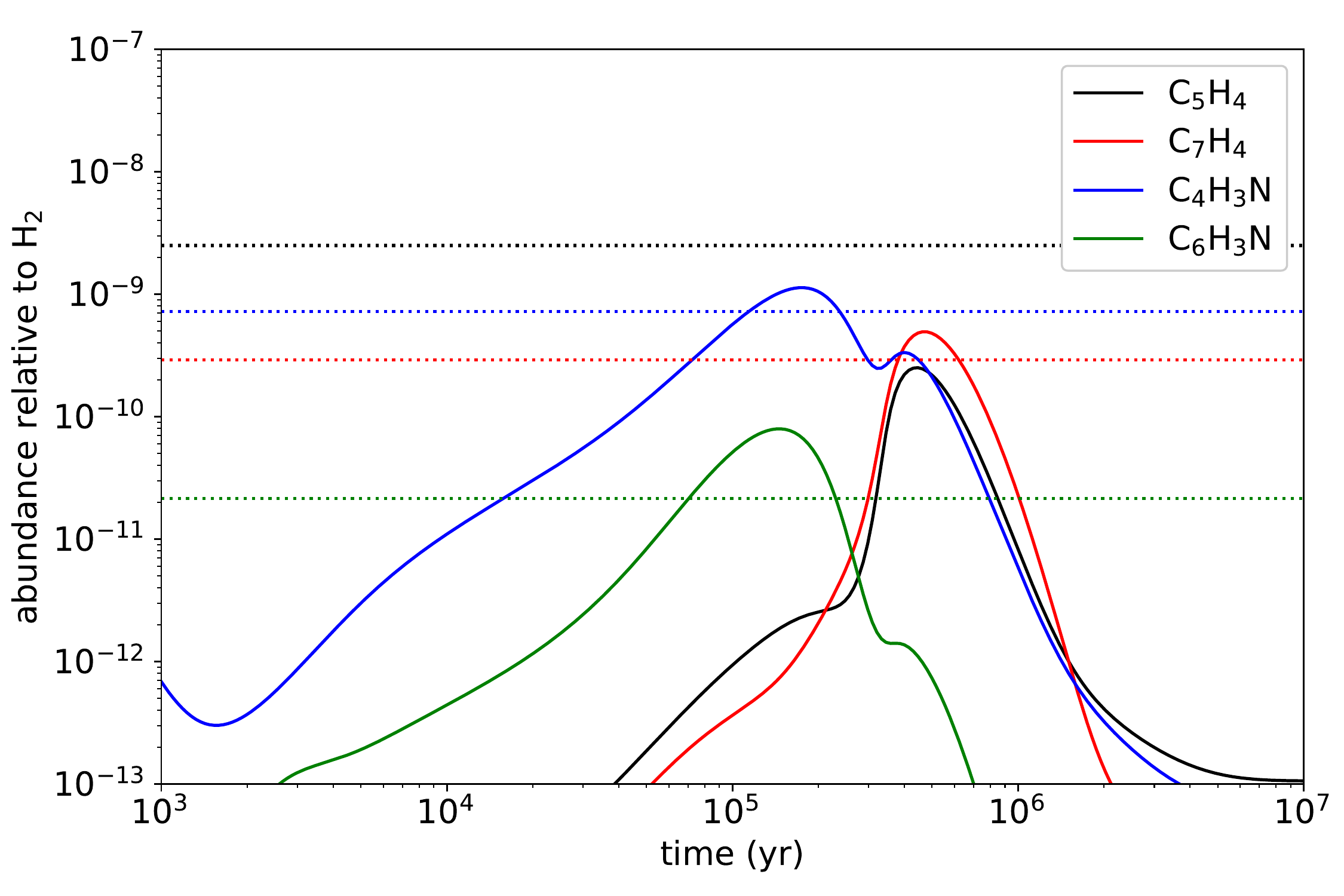}
\caption{Calculated fractional abundances of C$_5$H$_4$, C$_7$H$_4$, C$_4$H$_3$N, and C$_6$H$_3$N (allowing for various isomers) as a function of time. The horizontal dotted lines correspond to the abundances observed in \mbox{TMC-1}, where C$_5$H$_4$ includes CH$_3$C$_4$H and CH$_2$CCHCCH, C$_7$H$_4$ includes CH$_3$C$_6$H and CH$_2$CCHC$_4$H, C$_4$H$_3$N includes CH$_3$C$_3$N, CH$_2$CCHCN, and HCCCH$_2$CN, and C$_6$H$_3$N includes CH$_3$C$_5$N and CH$_2$CCHC$_3$N.}
\label{fig:abun}
\end{figure}

For a better understanding of the formation of CH$_2$CCHC$_4$H and CH$_2$CCHC$_3$N, we carried out chemical model calculations. The chemical model is similar to those presented in \cite{Marcelino2021} and \cite{Cernicharo2021e}. Briefly, we adopted typical parameters of cold dark clouds, namely, a gas kinetic temperature of 10 K, a volume density of H nuclei of 2\,$\times$\,10$^4$ cm$^{-3}$, a visual extinction of 30 mag, a cosmic-ray ionization rate of 1.3\,$\times$\,10$^{-17}$ s$^{-1}$, and the set of "low-metal" elemental abundances \citep{Agundez2013}. We adopted the chemical network RATE12 from the UMIST database \citep{McElroy2013}, updated to include the chemistry of C$_5$H$_4$, C$_7$H$_4$, C$_4$H$_3$N, and C$_6$H$_3$N. The chemistry of C$_5$H$_4$ and C$_4$H$_3$N isomers has been discussed by \cite{Cernicharo2021e} and \cite{Marcelino2021}, respectively. In the case of C$_7$H$_4$ isomers, the main formation routes involve the dissociative recombination of C$_7$H$_5^+$ with electrons and the neutral-neutral reactions of the C$_4$H radical with CH$_3$CCH, CH$_2$CCH$_2$, and CH$_3$CHCH$_2$, while C$_6$H$_3$N is formed by the dissociative recombination of C$_6$H$_4$N$^+$ and the neutral-neutral reactions of the radical C$_3$N with CH$_3$CCH and CH$_2$CCH$_2$. The reactions involving C$_4$H have been experimentally studied by \cite{Berteloite2010}, while those involving C$_3$N have been studied also experimentally by \cite{Fournier2014}. These reactions are found to be rapid at low temperatures, although information on the branching ratios of the different exothermic channels is not available. Therefore, in the chemical model, we do not distinguish between different isomers of C$_7$H$_4$ or C$_6$H$_3$N. The chemistry of these two isomeric families share similarities because an important part of the synthesis relies on the reaction of either C$_4$H or C$_3$N with methyl acetylene and allene. We may summarize these routes as:
\begin{subequations} \label{reac:c4h+ch3cch}
\begin{align}
\rm C_4H + CH_3CCH & \rightarrow \rm CH_3C_6H + H, \label{reac:c4h+ch3cch_a} \\
                                    & \rightarrow \rm CH_2CCHC_4H + H, \label{reac:c4h+ch3cch_b}
\end{align}
\end{subequations}
\begin{subequations} \label{reac:c4h+ch2cch2}
\begin{align}
\rm C_4H + CH_2CCH_2 & \rightarrow \rm CH_3C_6H + H, \label{reac:c4h+ch2cch2_a} \\
                                    & \rightarrow \rm CH_2CCHC_4H + H, \label{reac:c4h+ch2cch2_b}
\end{align}
\end{subequations}
\begin{subequations} \label{reac:c3n+ch3cch}
\begin{align}
\rm C_3N + CH_3CCH & \rightarrow \rm CH_3C_5N + H, \label{reac:c3n+ch3cch_a} \\
                                    & \rightarrow \rm CH_2CCHC_3N + H, \label{reac:c3n+ch3cch_b}
\end{align}
\end{subequations}
\begin{subequations} \label{reac:c3n+ch2cch2}
\begin{align}
\rm C_3N + CH_2CCH_2 & \rightarrow \rm CH_3C_5N + H, \label{reac:c3n+ch2cch2_a} \\
                                    & \rightarrow \rm CH_2CCHC_3N + H, \label{reac:c3n+ch2cch2_b}
\end{align}
\end{subequations}
where we only consider the C$_7$H$_4$ isomers CH$_3$C$_6$H and CH$_2$CCHC$_4$H because they are the only ones detected in \mbox{TMC-1} although other C$_7$H$_4$ isomers could also be formed (see \citealt{Berteloite2010}). Similarly, only the C$_6$H$_3$N isomers CH$_3$C$_5$N and CH$_2$CCHC$_3$N are considered, but others can also be formed. We adopt as main destruction routes for C$_7$H$_4$ and C$_6$H$_3$N the reactions with the ions C$^+$, He$^+$, HCO$^+$, and H$_3^+$. In Fig.~\ref{fig:abun} we compare the abundances calculated for C$_7$H$_4$ and C$_6$H$_3$N with the values observed in \mbox{TMC-1}. It is seen that peak calculated abundances, reached between 10$^5$ yr and 10$^6$ yr, are, within one order of magnitude, in good agreement with the observed values.

We note that the peak abundance is reached at different times for C$_7$H$_4$ and C$_6$H$_3$N. This is due to the different contribution of the neutral and ionic routes, both of which contribute to the formation of each species. The neutral route represented by reactions~(\ref{reac:c4h+ch3cch}-\ref{reac:c3n+ch2cch2}) is most efficient at times around (4-5)\,$\times$\,10$^5$ yr and this is the dominant route for C$_7$H$_4$. On the other hand, the ionic route, which involves as final step the dissociative recombination of C$_7$H$_5^+$ (to yield C$_7$H$_4$) and C$_6$H$_4$N$^+$ (to form C$_6$H$_3$N), is efficient at earlier times, in the range (1-2)\,$\times$\,10$^5$ yr, and this pathway becomes dominant over the neutral route when forming C$_6$H$_3$N.

Our chemical model is similar to that of \cite{Shingledecker2021} in that both explain the abundance derived for CH$_2$CCHC$_3$N in \mbox{TMC-1} through gas-phase chemistry. There are however some differences. In our case two gas-phase routes contribute to the formation of CH$_2$CCHC$_3$N: 1) the neutral one based on the reaction of C$_3$N with CH$_3$CCH and CH$_2$CCH$_2$ and 2) the ionic one based on the dissociative recombination of C$_6$H$_4$N$^+$, and the resulting peak abundance of CH$_2$CCHC$_3$N is in the range between 10$^{-11}$ and 10$^{-10}$ relative to H$_2$. In the model of \cite{Shingledecker2021}, CH$_2$CCHC$_3$N is mainly formed through the gas-phase neutral route and the peak abundance is substantially higher, in the range between 10$^{-9}$ and 10$^{-8}$ relative to H$_2$, probably because these authors consider a C/O ratio higher than 1, whereas we adopted a C/O ratio of 0.55.

The column densities of CH$_3$C$_6$H and CH$_3$C$_5$N in \mbox{TMC-1} are 7\,$\times$\,10$^{11}$ cm$^{-2}$ and 9.5\,$\times$\,10$^{10}$ cm$^{-2}$, respectively (see Appendix \ref{ch3c6h-5n}). This implies that the abundance ratio CH$_3$C$_6$H/CH$_2$CCHC$_4$H is 0.32, which is somewhat lower than that of the smaller analogue system CH$_3$C$_4$H/CH$_2$CCHCCH  = 1.1 \citep{Cernicharo2021e}. On the other hand, the abundance ratio CH$_3$C$_5$N/CH$_2$CCHC$_3$N is 0.79, which is similar to the abundance ratio of the smaller analogue system CH$_3$C$_3$N/CH$_2$CCHCN = 0.64 \citep{Marcelino2021}. If most of the synthesis of C$_7$H$_4$ and C$_6$H$_3$N isomers is carried out by reactions (1-4), this implies that CH$_2$CCHC$_4$H would be favoured over CH$_3$C$_6$H in reactions (1-2), while CH$_3$C$_5$N and CH$_2$CCHC$_3$N would be produced with similar branching ratios in reactions (3-4). Moreover, given that in the analogue smaller system C$_4$H$_3$N, three different isomers (CH$_3$C$_3$N, CH$_2$CCHCN, and HCCCH$_2$CN) have been detected in \mbox{TMC-1} with similar abundances \citep{Marcelino2021}, it would not be surprising to find that the isomer HCCCH$_2$C$_3$N would also be detected in \mbox{TMC-1}.

\section{Conclusions}
We report the first detection in space of CH$_2$CCHC$_4$H, towards the cold dark cloud TMC-1. The CH$_2$CCHC$_3$N molecule was also detected in a robust way, which allows for its frequencies
and rotational constants to be presented for the first time. We observed a total of 20 rotational transitions, from J=17 to 24 and $K_a\le 3$, using the Yebes 40m telescope. The frequencies predicted for these two species from laboratory data are rather accurate when compared with those derived from astronomical observations. The rotational temperature for both species is 9.0$\pm$0.5 K and the derived column densities are (2.2$\pm$0.2)$\times$10$^{12}$ cm$^{-2}$ for CH$_2$CCHC$_4$H and (1.2$\pm$0.15)$\times$10$^{11}$cm$^{-2}$ for CH$_2$CCHC$_3$N. These values give an abundance ratio between
CH$_2$CCHC$_4$H and CH$_2$CCHC$_3$N of 18$\pm$4. The observed abundances of the C$_7$H$_4$ and C$_6$H$_3$N isomeric families are reasonably well reproduced by a gas-phase chemical model in which the main formation pathways involve dissociative recombination of the precursor ions C$_7$H$_5^+$ and C$_6$H$_4^+$ and neutral-neutral reactions involving the C$_4$H and CN reactions with CH$_3$CCH and CH$_2$CCH$_2$.

\begin{acknowledgements}

We thank Ministerio de Ciencia e Innovaci\'on of Spain (MICIU) for funding support through projects
PID2019-106110GB-I00, PID2019-107115GB-C21 / AEI / 10.13039/501100011033, and
PID2019-106235GB-I00. We also thank ERC for funding
through grant ERC-2013-Syg-610256-NANOCOSMOS. M.A. thanks MICIU for grant RyC-2014-16277.
\end{acknowledgements}

\normalsize

\clearpage
\onecolumn

\begin{appendix}
\section{Observed line parameters}
\label{line_parameters}
The line parameters for the different molecules studied in this work were obtained by fitting a Gaussian line
profile to the observed data. A window of $\pm$ 15 \kms\, around the v$_{LSR}$ of the source was
considered for each transition. The derived line parameters for all the molecular species 
studied in this work are given in Table \ref{obs_line_parameters}. The observed lines of CH$_2$CCHC$_4$H are
shown on Fig. \ref{fig_ch2cchc4h}. Selected lines of CH$_2$CCHC$_3$N are shown on Fig. \ref{fig_ch2cchc3n}. All
lines of both species are weak, with many of them blended with other features. Nevertheless, for most of these
blended lines, it is still possible to derive reliable line parameters. It is worth noting that at this level
of sensitivity any stacking procedure is not at all recommended due to the forest of weak unknown features. For CH$_2$CCHC$_3$N,
we explored lines with $K_a$=3, but only three of them were detected. These lines
involve high energy levels and exhibit
intrinsic line strengths weaker than those with $K_a$=0, 1, and 2. This species also demonstrate a hyperfine structure due to the 
nitrogen nuclear spin. None of the hyperfine structures are resolved as the splitting is predicted to be of $\sim$5 kHz for $K_a$=0 and 1 
and around 10 kHz for $K_a$=2; this is much lower than our spectral resolution of 38.1 kHz.
However, the observed linewidths for many of the lines of this species indicate some broadening. For the modelling of
the emerging spectrum of this molecule, we used predictions taken into account the hyperfine splitting.

\begin{longtable}{cccccl}
\caption{Observed line parameters for the species studied in  this work.} \label{obs_line_parameters}\\
\hline 
\hline
Transition$^1$     &$\nu_{rest}$~$^a$    & $\int T_A^* dv$~$^b$&  \multicolumn{1}{c}{$\Delta$v$^c$}    & \multicolumn{1}{c}{$T_A^*$$^d$} & Notes  \\
                   &  (MHz)              & (mK\,km\,s$^{-1}$)  &  \multicolumn{1}{c}{(km\,s$^{-1}$)}    & \multicolumn{1}{c}{(mK)}       &  \\
\hline
\endfirsthead
\caption{continued.}\\
\hline \hline
Transition$^1$     &$\nu_{obs}$~$^a$    & $\int T_A^* dv$~$^b$ & \multicolumn{1}{c}{$\Delta$v$^c$}    & \multicolumn{1}{c}{$T_A^*$$^d$} & Notes \\
                   &  (MHz)             & (mK\,km\,s$^{-1}$)   & \multicolumn{1}{c}{(km\,s$^{-1}$)}    & \multicolumn{1}{c}{(mK)}       &  \\
\hline
\endhead
\hline
\endfoot
\hline
\multicolumn{6}{c}{\bf CH$_2$CCHC$_4$H$^1$} \\
$17_{0,17}-16_{0,16}$& 31281.977$\pm$0.015& 0.36$\pm$0.10& 0.68$\pm$0.15&  0.50$\pm$0.12&  \\
$17_{2,16}-16_{2,15}$& 31369.493$\pm$0.002&              &              & $\le$0.45     & A\\
$17_{2,15}-16_{2,14}$& 31470.108$\pm$0.015& 0.35$\pm$0.11& 0.54$\pm$0.20&  0.60$\pm$0.17& B\\
$17_{1,16}-16_{1,15}$& 31778.167$\pm$0.030& 0.47$\pm$0.09& 0.66$\pm$0.00&  0.66$\pm$0.14& C\\
$18_{1,18}-17_{1,17}$& 32754.991$\pm$0.015& 0.59$\pm$0.16& 1.27$\pm$0.37&  0.44$\pm$0.14&  \\
$18_{0,18}-17_{0,17}$& 33109.190$\pm$0.002&              &              &               & A\\
$18_{2,17}-17_{2,16}$& 33212.230$\pm$0.020& 0.30$\pm$0.08& 0.83$\pm$0.24&  0.34$\pm$0.11& D\\
$18_{2,16}-17_{2,15}$& 33331.250$\pm$0.015& 0.54$\pm$0.15& 0.74$\pm$0.14&  0.68$\pm$0.11&  \\
$18_{1,17}-17_{1,16}$& 33643.419$\pm$0.015& 0.48$\pm$0.11& 0.64$\pm$0.16&  0.71$\pm$0.13&  \\
$19_{1,19}-18_{1,18}$& 34571.094$\pm$0.015& 0.40$\pm$0.08& 0.56$\pm$0.13&  0.67$\pm$0.14&  \\
$19_{0,19}-18_{0,18}$& 34934.337$\pm$0.015& 0.35$\pm$0.10& 0.54$\pm$0.19&  0.59$\pm$0.14&  \\
$19_{2,18}-18_{2,17}$& 35054.486$\pm$0.030& 0.41$\pm$0.10& 0.75$\pm$0.00&  0.50$\pm$0.14& E\\
$19_{2,17}-18_{2,16}$& 35193.985$\pm$0.004&              &              & $\le$0.44     & A\\
$19_{1,18}-18_{1,17}$& 35507.943$\pm$0.003&              &              &               & F\\ 
$20_{1,20}-19_{1,19}$& 36386.699$\pm$0.015& 0.26$\pm$0.09& 0.55$\pm$0.12&  0.45$\pm$0.11& \\
$20_{0,20}-19_{0,19}$& 36757.421$\pm$0.015& 0.64$\pm$0.13& 0.96$\pm$0.25&  0.63$\pm$0.14& \\
$20_{2,19}-19_{2,18}$& 36896.323$\pm$0.020& 0.25$\pm$0.08& 0.55$\pm$0.22&  0.48$\pm$0.14& B\\
$20_{2,18}-19_{2,17}$& 37058.291$\pm$0.015& 0.43$\pm$0.09& 0.59$\pm$0.16&  0.70$\pm$0.15& \\
$20_{1,19}-19_{1,18}$& 37371.695$\pm$0.004&              &              &  $\le$0.42    & A\\
$21_{1,21}-20_{1,20}$& 38201.714$\pm$0.015& 0.29$\pm$0.08& 0.72$\pm$0.23&  0.38$\pm$0.12& \\
$21_{0,21}-20_{0,20}$& 38578.311$\pm$0.015& 0.19$\pm$0.06& 0.37$\pm$0.14&  0.48$\pm$0.13& \\
$21_{2,20}-20_{2,19}$& 38737.658$\pm$0.015& 0.14$\pm$0.05& 0.29$\pm$0.15&  0.42$\pm$0.12& \\
$21_{2,19}-20_{2,18}$& 38924.186$\pm$0.015& 0.16$\pm$0.06& 0.30$\pm$0.17&  0.50$\pm$0.13& \\
$21_{1,20}-20_{1,19}$& 39234.610$\pm$0.005&              &              &  $\le$1.2     & A,G\\
$22_{1,22}-21_{1,21}$& 40016.225$\pm$0.015& 0.38$\pm$0.09& 0.49$\pm$0.15&  0.73$\pm$0.15& \\
$22_{0,22}-21_{0,21}$& 40397.015$\pm$0.015& 0.29$\pm$0.08& 0.64$\pm$0.27&  0.42$\pm$0.16& \\
$22_{1,21}-21_{1,20}$& 41096.635$\pm$0.006&              &              &  $\le$0.55    & A\\     
$23_{1,23}-22_{1,22}$& 41830.115$\pm$0.006&              &              &               & H\\
$23_{0,23}-22_{0,22}$& 42213.561$\pm$0.004&              &              &  $\le$0.48    & A\\     
$23_{1,22}-22_{1,21}$& 42957.710$\pm$0.015& 0.34$\pm$0.12& 0.61$\pm$0.20&  0.51$\pm$0.18&  \\
$24_{0,24}-23_{0,23}$& 44027.907$\pm$0.005&              &              &   $\le$0.60   & A,I\\
\hline
\multicolumn{6}{c}{\bf CH$_2$CCHC$_3$N$^1$} \\
$17_{1,17}-16_{1,16}$& 31112.574$\pm$0.015& 0.75$\pm$0.15& 0.77$\pm$0.00&  0.92$\pm$0.12& C\\
$17_{0,17}-16_{0,16}$& 31463.463$\pm$0.015& 0.99$\pm$0.21& 1.17$\pm$0.35&  0.80$\pm$0.12&  \\
$17_{2,16}-16_{2,15}$& 31558.376$\pm$0.015& 0.44$\pm$0.12& 0.52$\pm$0.18&  0.76$\pm$0.18&  \\
$17_{3,15}-16_{3,14}$& 31593.484$\pm$0.020& 0.38$\pm$0.11& 0.65$\pm$0.19&  0.55$\pm$0.18&  \\
$17_{3,14}-16_{3,13}$& 31596.159$\pm$0.016&              &              & $\le$0.54     &  \\
$17_{2,15}-16_{2,14}$& 31667.997$\pm$0.015& 0.44$\pm$0.11& 0.73$\pm$0.18&  0.56$\pm$0.14&  \\
$17_{1,16}-16_{1,15}$& 31980.224$\pm$0.010& 0.86$\pm$0.09& 0.81$\pm$0.10&  0.99$\pm$0.13&  \\
$18_{1,18}-17_{1,17}$& 32939.203$\pm$0.010& 0.57$\pm$0.08& 0.78$\pm$0.12&  0.68$\pm$0.12&  \\
$18_{0,18}-17_{0,17}$& 33300.280$\pm$0.010& 0.97$\pm$0.11& 0.92$\pm$0.13&  0.99$\pm$0.13&  \\
$18_{2,17}-17_{2,16}$& 33411.994$\pm$0.015& 0.52$\pm$0.16& 1.12$\pm$0.29&  0.51$\pm$0.13&  \\
$18_{3,16}-17_{3,15}$& 33453.015$\pm$0.017&              &              & $\le$0.40     & I\\
$18_{3,15}-17_{3,14}$& 33456.519$\pm$0.016&              &              &               & D\\
$18_{2,16}-17_{2,15}$& 33541.670$\pm$0.015& 0.53$\pm$0.12& 0.88$\pm$0.27&  0.57$\pm$0.13& \\
$18_{1,17}-17_{1,16}$& 33857.014$\pm$0.010& 0.59$\pm$0.09& 0.78$\pm$0.13&  0.72$\pm$0.12&  \\
$19_{1,19}-18_{1,18}$& 34765.294$\pm$0.015& 0.53$\pm$0.10& 0.69$\pm$0.14&  0.72$\pm$0.14&  \\
$19_{0,19}-18_{0,18}$& 35134.847$\pm$0.020& 0.35$\pm$0.11& 0.50$\pm$0.16&  0.66$\pm$0.16&  \\
$19_{2,18}-18_{2,17}$& 35265.143$\pm$0.020& 0.54$\pm$0.06& 0.60$\pm$0.00&  0.85$\pm$0.12& C\\
$19_{3,17}-18_{3,16}$& 35312.608$\pm$0.015& 0.36$\pm$0.10& 0.65$\pm$0.20&  0.51$\pm$0.15& \\
$19_{3,16}-18_{3,15}$& 35317.258$\pm$0.020& 0.21$\pm$0.08& 0.67$\pm$0.23&  0.34$\pm$0.14& \\ 
$19_{2,17}-18_{2,16}$& 35416.968$\pm$0.020& 0.24$\pm$0.07& 0.61$\pm$0.24&  0.36$\pm$0.11&  \\
$19_{1,18}-18_{1,17}$& 35733.011$\pm$0.020& 0.41$\pm$0.13& 0.63$\pm$0.25&  0.61$\pm$0.17&  \\
$20_{1,20}-19_{1,19}$& 36590.744$\pm$0.010& 0.55$\pm$0.11& 0.90$\pm$0.17&  0.57$\pm$0.14& \\
$20_{0,20}-19_{0,19}$& 36967.205$\pm$0.015& 0.27$\pm$0.07& 0.61$\pm$0.16&  0.42$\pm$0.12&  \\
$20_{2,19}-19_{2,18}$& 37117.811$\pm$0.010& 0.44$\pm$0.08& 0.65$\pm$0.11&  0.64$\pm$0.12&  \\
$20_{2,18}-19_{2,17}$& 37293.956$\pm$0.020& 0.55$\pm$0.19& 0.93$\pm$0.33&  0.56$\pm$0.13& B\\
$21_{2,20}-20_{2,19}$& 38969.920$\pm$0.007&              &              & $\le$0.42     & A\\
$21_{2,19}-20_{2,18}$& 39172.697$\pm$0.008&              &              & $\le$0.45     & A\\
$22_{2,21}-21_{2,20}$& 40821.522$\pm$0.010& 0.47$\pm$0.12& 0.65$\pm$0.18&  0.69$\pm$0.18& \\
$22_{2,20}-21_{2,19}$& 41053.129$\pm$0.020& 0.28$\pm$0.07& 0.60$\pm$0.16&  0.44$\pm$0.12& \\
$22_{1,21}-21_{1,20}$& 41355.602$\pm$0.020& 0.20$\pm$0.08& 0.38$\pm$0.13&  0.66$\pm$0.18& \\
$23_{1,23}-22_{1,22}$& 42063.663$\pm$0.015& 0.35$\pm$0.08& 0.61$\pm$0.19&  0.53$\pm$0.12& \\
$23_{0,23}-22_{0,22}$& 42450.325$\pm$0.020& 0.38$\pm$0.09& 0.54$\pm$0.18&  0.65$\pm$0.18& \\
$23_{2,22}-22_{2,21}$& 42672.553$\pm$0.020& 0.21$\pm$0.08& 0.40$\pm$0.24&  0.49$\pm$0.18& \\
$23_{2,21}-22_{2,20}$&                    &              &              & $\le$0.51     & A\\
$23_{1,22}-22_{1,21}$& 43227.824$\pm$0.005&              &              &               & D\\
\hline
\end{longtable}
\tablefoot{\\
\tablefoottext{1}{Quantum numbers are $J'_{K'_{a,}K'_{c}}$ - $J_{K_{a,}K_{c}}$.}\\
\tablefoottext{a}{Observed frequency of the transition assuming a LSR velocity of 5.83 km s$^{-1}$. See 
text for the laboratory data used for initial frequency predictions of each molecular species. For lines 
for which only upper limits are given, their frequencies
correspond to the predictions obtained from a fit to the laboratory plus space lines 
(see Table \ref{new_rotational_ch2cchc4h}).}\\
\tablefoottext{b}{Integrated line intensity in mK\,km\,s$^{-1}$.}\\
\tablefoottext{c}{Linewidth at half intensity derived by fitting a Gaussian function to
the observed line profile (in km\,s$^{-1}$).}\\
\tablefoottext{d}{Antenna temperature in milli Kelvin.}\\
\tablefoottext{A}{Upper limit corresponds to 3$\sigma$.}\\
\tablefoottext{B}{The line is partially blended but still possible to derive line parameters.}\\
\tablefoottext{C}{This line is blended with an U line. Line width has been fixed.}\\
\tablefoottext{D}{This line is blended with a feature stronger than expected. It is, more likely, an unknown line.}\\
\tablefoottext{E}{This line is blended with a transition of H$_2$C$_4$N.}\\
\tablefoottext{F}{This line is heavily blended a transition of C$^{13}$CCCCH. Line parameters can not
be derived.}\\
\tablefoottext{G}{The frequency of this transition of CH$_2$CCHC$_4$H is within a poor sensitivity zone 
of the receiver.}\\
\tablefoottext{H}{This line is blended with a broad 1 mK unknown line.}\\
\tablefoottext{I}{A 3$\sigma$ feature do appears just at the expected frequency of this transition.}\\
}
\twocolumn

\onecolumn
\section{Improved rotational constants for CH$_2$CCHC$_4$H and CH$_2$CCHC$_3$N}
\label{new_constants}
We merged the laboratory lines of CH$_2$CCHC$_4$H and CH$_2$CCHC$_3$N with
those observed in TMC-1 with the QUIJOTE line survey to obtain improved
rotational and distortion constants for both species. The results for
CH$_2$CCHC$_4$H are given in Table \ref{new_rotational_ch2cchc4h}. The
predictions from the laboratory data were rather accurate, with deviations
between our observed frequencies of a few kHz (see  Table \ref{obs_frequencies_ch2cchc4h}). Some lines show differences
of up to 25 kHz which is at most 1.5-2 times the accuracy of the observed
frequencies in TMC-1 as the signal-to-noise ratio (S/N) of the observations is
limited to $\sim$4-5. As a consequence, the derived constants for this
species from the merged fit are practically identical to those obtained
from the laboratory data alone. However, the uncertainty of the constants
is significantly improved.

The results for CH$_2$CCHC$_3$N are given in Table \ref{new_rotational_ch2cchc3n}.
The new constants are very similar to the laboratory data, with most of the differences being at frequencies below 3 kHz (see Table \ref{obs_frequencies_ch2cchc3n}) and presenting a maximum deviation of 47 kHz. As in the previous case, the values of the constants obtained are in good agreement with the laboratory data.

\begin{table}[h]
\caption{Improved rotational and distortion constants for CH$_2$CCHC$_4$H.}
\label{new_rotational_ch2cchc4h}
\centering
\begin{tabular}{{|c|c|c|}}
\hline
Constant                 & Laboratory$^1$ & Lab + TMC-1$^2$ \\
(MHz)                    &                &                 \\
\hline
$A  $                    &15654.01571(87) & 15654.01571(97) \\
$B  $                    &  947.90683(11) &   947.906800(92)\\
$C  $                    &  898.29939(10) &   898.299386(82)\\
$D_J$\,$\times$10$^4$    &    1.4692(15)  &     1.4686(10)  \\
$D_{JK}$\,$\times$10$^2$ &   -2.48883(88) &    -2.48885(93) \\
$d_J$\,$\times$10$^5$    &    3.100(20)   &     3.097(14)   \\
\hline
\hline
$J_{max}$                &     13         &        23       \\    
$K_{a,max}$               &      2         &        2         \\
$N_{lines}$              &      45        &        67         \\
$\sigma$ (kHz)            &      1.7       &        9.3        \\
$\nu_{max}$ (MHz)        &    24310       &      42957        \\
\hline
\end{tabular}
\tablefoot{\\
        \tablefoottext{1}{Rotational and distortion constants derived from a
       fit to the laboratory data of \citet{McCarthy2020}.}\\
        \tablefoottext{2}{Merged rotational and distortion constants from a fit to
       the laboratory and TMC-1 frequencies.}\\
}
\end{table}
\normalsize

\begin{table}[h]
\caption{Improved rotational and distortion constants for CH$_2$CCHC$_3$N.}
\label{new_rotational_ch2cchc3n}
\centering
\begin{tabular}{{|c|c|c|}}
\hline
Constant                 & Laboratory$^1$ & Lab + TMC-1$^2$ \\
(MHz)                    &                &                 \\
\hline
$A  $                    &   15351.89314(36)  &   15351.89321(41)    \\
$B  $                    &    954.343246(105) &    954.343014(70)    \\
$C  $                    &    903.075581(88)  &    903.075751(62)    \\
$D_J$\,$\times$10$^4$    &     1.53425(187)   &     1.53243(73)      \\
$D_{JK}$\,$\times$10$^2$ &    -24.1218(75)    &     -24.1204(78)     \\
$d_J$\,$\times$10$^5$    &     3.2427(274)    &      3.1674(92)      \\
$\chi_{aa}$              &    -3.77238(147)   &      -3.77234(167)   \\
$\chi_{bb}$              &     1.74588(148)   &       1.74567(168)   \\
\hline
\hline
$J_{max}$                &  15  &     23 \\
$K_{a,max}$              &  2   &      3  \\
$N_{lines}$              &  100 &      134  \\
$\sigma$ (kHz)            &   1.7   &    8.9    \\
$\nu_{max}$ (MHz)        &   27783   &    43227    \\
\hline
\end{tabular}
\tablefoot{\\
        \tablefoottext{1}{Rotational and distortion constants kindly provided by M. McCarthy.}\\
        \tablefoottext{2}{Merged rotational and distortion constants from a fit to the laboratory and TMC-1 frequencies.}\\
}
\end{table}
\normalsize

\onecolumn
\begin{table}
\caption{Observed frequencies of CH$_2$CCHC$_4$H}
\label{obs_frequencies_ch2cchc4h}
z\centering
\small
\begin{tabular}{|l|r|r|r|}
\hline
Transition              & $\nu_{obs}$         &$\nu_{obs}-\nu_{cal}$& Ref \\
$J'_{K'_{a,}K'_{c}}$ - $J_{K_{a,}K_{c}}$        &   (MHz)             &    (kHz)  &  \\        
\hline
$ 4_{1, 4}- 3_{1, 3}$   &  7285.5300$\pm$0.002&    0.60& 1\\        
$ 4_{0, 4}- 3_{0, 3}$   &  7383.5340$\pm$0.002&   -0.48& 1\\        
$ 4_{1, 3}- 3_{1, 2}$   &  7483.9430$\pm$0.002&    0.98& 1\\        
$ 5_{1, 5}- 4_{1, 4}$   &  9106.6580$\pm$0.002&    0.61& 1\\        
$ 5_{0, 5}- 4_{0, 4}$   &  9228.4530$\pm$0.002&    0.47& 1\\        
$ 5_{1, 4}- 4_{1, 3}$   &  9354.6610$\pm$0.002&    1.66& 1\\        
$ 6_{1, 6}- 5_{1, 5}$   & 10927.6180$\pm$0.002&    1.05& 1\\        
$ 6_{0, 6}- 5_{0, 5}$   & 11072.7280$\pm$0.002&    0.53& 1\\        
$ 6_{1, 5}- 5_{1, 4}$   & 11225.1980$\pm$0.002&    1.15& 1\\        
$ 7_{1, 7}- 6_{1, 6}$   & 12748.3760$\pm$0.002&    0.46& 1\\        
$ 7_{0, 7}- 6_{0, 6}$   & 12916.2310$\pm$0.002&   -0.52& 1\\        
$ 7_{1, 6}- 6_{1, 5}$   & 13095.5180$\pm$0.002&    0.44& 1\\        
$ 8_{1, 8}- 7_{1, 7}$   & 14568.9020$\pm$0.002&    0.73& 1\\        
$ 8_{0, 8}- 7_{0, 7}$   & 14758.8370$\pm$0.002&   -0.64& 1\\        
$ 8_{1, 7}- 7_{1, 6}$   & 14965.5850$\pm$0.002&    1.06& 1\\        
$ 9_{1, 9}- 8_{1, 8}$   & 16389.1630$\pm$0.002&    0.08& 1\\        
$ 9_{0, 9}- 8_{0, 8}$   & 16600.4190$\pm$0.002&   -0.92& 1\\        
$ 9_{1, 8}- 8_{1, 7}$   & 16835.3570$\pm$0.002&   -0.77& 1\\        
$10_{1,10}- 9_{1, 9}$   & 18209.1310$\pm$0.002&    0.88& 1\\        
$10_{0,10}- 9_{0, 9}$   & 18440.8540$\pm$0.002&   -0.02& 1\\        
$11_{0,11}-10_{0,10}$   & 20280.0170$\pm$0.002&   -0.81& 1\\        
$11_{1,10}-10_{1, 9}$   & 20573.8720$\pm$0.002&    0.81& 1\\        
$12_{1,12}-11_{1,11}$   & 21848.0660$\pm$0.002&    1.66& 1\\        
$12_{0,12}-11_{0,11}$   & 22117.7910$\pm$0.002&   -1.06& 1\\        
$12_{1,11}-11_{1,10}$   & 22442.5300$\pm$0.002&   -0.23& 1\\        
$13_{1,13}-12_{1,12}$   & 23666.9750$\pm$0.002&   -0.53& 1\\        
$13_{0,13}-12_{0,12}$   & 23954.0590$\pm$0.002&   -2.32& 1\\        
$13_{1,12}-12_{1,11}$   & 24310.7360$\pm$0.002&    0.52& 1\\        
$ 5_{2, 4}- 4_{2, 3}$   &  9231.6750$\pm$0.002&    0.32& 1\\        
$ 5_{2, 3}- 4_{2, 2}$   &  9234.1780$\pm$0.002&   -1.65& 1\\        
$ 7_{2, 6}- 6_{2, 5}$   & 12923.6610$\pm$0.002&   -0.54& 1\\        
$ 6_{2, 5}- 5_{2, 4}$   & 11077.7410$\pm$0.002&   -0.25& 1\\        
$ 6_{2, 4}- 5_{2, 3}$   & 11082.1220$\pm$0.002&   -2.00& 1\\        
$ 7_{2, 5}- 6_{2, 4}$   & 12930.6710$\pm$0.002&   -0.79& 1\\        
$ 8_{2, 7}- 7_{2, 6}$   & 14769.4100$\pm$0.002&   -1.21& 1\\        
$ 8_{2, 6}- 7_{2, 5}$   & 14779.9210$\pm$0.002&   -1.20& 1\\        
$ 9_{2, 8}- 8_{2, 7}$   & 16614.9660$\pm$0.002&    0.00& 1\\        
$ 9_{2, 7}- 8_{2, 6}$   & 16629.9720$\pm$0.002&   -1.31& 1\\        
$10_{2, 9}- 9_{2, 8}$   & 18460.3080$\pm$0.002&    6.34& 1\\        
$10_{2, 8}- 9_{2, 7}$   & 18480.9200$\pm$0.002&   -1.63& 1\\        
$ 1_{1, 0}- 1_{0, 1}$   & 14755.7680$\pm$0.002&    2.02& 1\\        
$ 1_{1, 1}- 0_{0, 0}$   & 16552.3630$\pm$0.002&   -1.41& 1\\        
$ 2_{1, 2}- 1_{0, 1}$   & 18349.0590$\pm$0.002&   -0.61& 1\\        
$11_{2,10}-10_{2, 9}$   & 20305.3950$\pm$0.002&    0.97& 1\\        
$11_{2, 9}-10_{2, 8}$   & 20332.8600$\pm$0.002&   -1.50& 1\\        
$17_{0,17}-16_{0,16}$   & 31281.9770$\pm$0.015&    7.13& 2\\          
$17_{2,15}-16_{2,14}$   & 31470.1080$\pm$0.015&  -14.60& 2\\          
$17_{1,16}-16_{1,15}$   & 31778.1670$\pm$0.030&   25.67& 2\\          
$18_{1,18}-17_{1,17}$   & 32754.9910$\pm$0.015&   26.74& 2\\          
$18_{2,17}-17_{2,16}$   & 33212.2300$\pm$0.020&   18.16& 2\\          
$18_{2,16}-17_{2,15}$   & 33331.2500$\pm$0.015&  -27.35& 2\\          
$18_{1,17}-17_{1,16}$   & 33643.4190$\pm$0.015&   11.01& 2\\          
$19_{1,19}-18_{1,18}$   & 34571.0940$\pm$0.015&    0.50& 2\\          
$19_{0,19}-18_{0,18}$   & 34934.3370$\pm$0.015&  -17.72& 2\\          
$19_{2,18}-18_{2,17}$   & 35054.4860$\pm$0.030&  -10.68& 2\\          
$20_{1,20}-19_{1,19}$   & 36386.6990$\pm$0.015&   10.54& 2\\          
$20_{0,20}-19_{0,19}$   & 36757.4210$\pm$0.015&   12.62& 2\\          
$20_{2,19}-19_{2,18}$   & 36896.3230$\pm$0.020&  -01.70& 2\\          
$20_{2,18}-19_{2,17}$   & 37058.2910$\pm$0.015&   13.42& 2\\          
$21_{1,21}-20_{1,20}$   & 38201.7140$\pm$0.015&  -19.11& 2\\          
$21_{0,21}-20_{0,20}$   & 38578.3110$\pm$0.015&    1.88& 2\\          
$21_{2,20}-20_{2,19}$   & 38737.6580$\pm$0.015&  -14.91& 2\\          
$21_{2,19}-20_{2,18}$   & 38924.1860$\pm$0.015&   21.35& 2\\          
$22_{1,22}-21_{1,21}$   & 40016.2250$\pm$0.015&   11.95& 2\\          
$22_{0,22}-21_{0,21}$   & 40397.0150$\pm$0.015&  -15.16& 2\\          
$23_{1,22}-22_{1,21}$   & 42957.7100$\pm$0.015&    0.24& 2\\          
\hline
\end{tabular}
\tablefoot{\\
        \tablefoottext{1}{Laboratory frequencies from \citet{McCarthy2020}.}\\
        \tablefoottext{2}{TMC-1 frequencies from Table \ref{obs_line_parameters}.}\\
}
\end{table}
\normalsize

\onecolumn
\begin{longtable}{|l|r|r|r|r|r|}
\caption{Observed frequencies of CH$_2$CCHC$_3$N} \label{obs_frequencies_ch2cchc3n}\\
\hline 
\hline
Transition     & $F'$ & $F$            & $\nu_{obs}$         &$\nu_{obs}-\nu_{cal}$& Ref \\
 $J'_{K'_{a,}K'_{c}}$ - $J_{K_{a,}K_{c}}$              &      &                  &   (MHz)             &    (kHz)  &  \\
\hline
\endfirsthead
\caption{continued.}\\
\hline \hline
Transition     & $F'$ & $F$            & $\nu_{obs}$         &$\nu_{obs}-\nu_{cal}$& Ref \\
$J'_{K'_{a,}K'_{c}}$ - $J_{K_{a,}K_{c}}$               &      &                  &   (MHz)             &    (kHz)  &  \\
\hline
\endhead
\hline
\endfoot
\hline
 $  4_{0, 4}-3_{0, 3}  $ &  4 & 4  &    7427.010$\pm$0.002  &  -1.41  & 1  \\
 $  4_{0, 4}-3_{0, 3}  $ &  3 & 2  &    7428.189$\pm$0.002  &   0.36  & 1  \\
 $  4_{0, 4}-3_{0, 3}  $ &  4 & 3  &    7428.270$\pm$0.002  &   0.86  & 1  \\
 $  4_{0, 4}-3_{0, 3}  $ &  5 & 4  &    7428.298$\pm$0.002  &  -0.08  & 1  \\
 $  4_{0, 4}-3_{0, 3}  $ &  3 & 3  &    7429.886$\pm$0.002  &  -0.94  & 1  \\
 $  6_{0, 6}-5_{0, 5}  $ &  5 & 4  &   11139.572$\pm$0.002  &   0.94  & 1  \\
 $  6_{0, 6}-5_{0, 5}  $ &  6 & 5  &   11139.598$\pm$0.002  &  -1.09  & 1  \\
 $  6_{0, 6}-5_{0, 5}  $ &  7 & 6  &   11139.615$\pm$0.002  &   0.87  & 1  \\
 $  7_{1, 7}-6_{1, 6}  $ &  7 & 6  &   12820.872$\pm$0.002  &   0.78  & 1  \\
 $  7_{1, 7}-6_{1, 6}  $ &  6 & 5  &   12820.882$\pm$0.002  &   2.54  & 1  \\
 $  7_{1, 7}-6_{1, 6}  $ &  8 & 7  &   12820.907$\pm$0.002  &   1.14  & 1  \\
 $  7_{0, 7}-6_{0, 6}  $ &  6 & 5  &   12994.054$\pm$0.002  &  -0.82  & 1  \\
 $  7_{0, 7}-6_{0, 6}  $ &  7 & 6  &   12994.073$\pm$0.002  &  -0.97  & 1  \\
 $  7_{0, 7}-6_{0, 6}  $ &  8 & 7  &   12994.086$\pm$0.002  &   0.32  & 1  \\
 $  7_{1, 6}-6_{1, 5}  $ &  7 & 6  &   13179.626$\pm$0.002  &  -0.47  & 1  \\
 $  7_{1, 6}-6_{1, 5}  $ &  6 & 5  &   13179.638$\pm$0.002  &   1.81  & 1  \\
 $  7_{1, 6}-6_{1, 5}  $ &  8 & 7  &   13179.660$\pm$0.002  &  -0.30  & 1  \\
 $  1_{1, 0}-1_{0, 1}  $ &  1 & 0  &   14447.486$\pm$0.002  &   0.45  & 1  \\
 $  1_{1, 0}-1_{0, 1}  $ &  2 & 2  &   14448.578$\pm$0.002  &   2.41  & 1  \\
 $  1_{1, 0}-1_{0, 1}  $ &  0 & 1  &   14448.796$\pm$0.002  &   0.68  & 1  \\
 $  1_{1, 0}-1_{0, 1}  $ &  1 & 2  &   14449.184$\pm$0.002  &   0.69  & 1  \\
 $  1_{1, 0}-1_{0, 1}  $ &  2 & 1  &   14449.709$\pm$0.002  &   1.92  & 1  \\
 $  1_{1, 0}-1_{0, 1}  $ &  1 & 1  &   14450.314$\pm$0.002  &  -0.80  & 1  \\
 $  2_{1, 1}-2_{0, 2}  $ &  2 & 1  &   14498.986$\pm$0.002  &   0.98  & 1  \\
 $  2_{1, 1}-2_{0, 2}  $ &  1 & 1  &   14499.859$\pm$0.002  &   0.58  & 1  \\
 $  2_{1, 1}-2_{0, 2}  $ &  3 & 3  &   14500.222$\pm$0.002  &   1.84  & 1  \\
 $  2_{1, 1}-2_{0, 2}  $ &  2 & 2  &   14500.872$\pm$0.002  &   0.25  & 1  \\
 $  2_{1, 1}-2_{0, 2}  $ &  1 & 2  &   14501.745$\pm$0.002  &  -0.15  & 1  \\
 $  3_{1, 2}-3_{0, 3}  $ &  3 & 4  &   14576.883$\pm$0.002  &   1.29  & 1  \\
 $  3_{1, 2}-3_{0, 3}  $ &  4 & 4  &   14577.779$\pm$0.002  &   1.12  & 1  \\
 $  3_{1, 2}-3_{0, 3}  $ &  3 & 3  &   14578.140$\pm$0.002  &   0.56  & 1  \\
 $  8_{0, 8}-7_{0, 7}  $ &  7 & 6  &   14847.557$\pm$0.002  &  -0.02  & 1  \\
 $  8_{0, 8}-7_{0, 7}  $ &  8 & 7  &   14847.570$\pm$0.002  &  -0.80  & 1  \\
 $  8_{0, 8}-7_{0, 7}  $ &  9 & 8  &   14847.582$\pm$0.002  &   1.74  & 1  \\
 $  8_{2, 7}-7_{2, 6}  $ &  8 & 7  &   14858.869$\pm$0.002  &   3.38  & 1  \\
 $  8_{2, 7}-7_{2, 6}  $ &  7 & 6  &   14858.926$\pm$0.002  &  -0.18  & 1  \\
 $  8_{2, 7}-7_{2, 6}  $ &  9 & 8  &   14858.939$\pm$0.002  &   2.56  & 1  \\
 $  8_{2, 6}-7_{2, 5}  $ &  8 & 7  &   14870.330$\pm$0.002  &  -0.99  & 1  \\
 $  8_{2, 6}-7_{2, 5}  $ &  7 & 6  &   14870.386$\pm$0.002  &  -4.81  & 1  \\
 $  8_{2, 6}-7_{2, 5}  $ &  9 & 8  &   14870.398$\pm$0.002  &  -3.11  & 1  \\
 $  8_{1, 7}-7_{1, 6}  $ &  8 & 7  &   15061.654$\pm$0.002  &  -0.55  & 1  \\
 $  8_{1, 7}-7_{1, 6}  $ &  7 & 6  &   15061.661$\pm$0.002  &   1.50  & 1  \\
 $  8_{1, 7}-7_{1, 6}  $ &  9 & 8  &   15061.678$\pm$0.002  &  -0.66  & 1  \\
 $  1_{1, 1}-0_{0, 0}  $ &  0 & 1  &   16254.142$\pm$0.002  &  -2.20  & 1  \\
 $  1_{1, 1}-0_{0, 0}  $ &  2 & 1  &   16254.929$\pm$0.002  &  -0.51  & 1  \\
 $  1_{1, 1}-0_{0, 0}  $ &  1 & 1  &   16255.451$\pm$0.002  &  -1.90  & 1  \\
 $  9_{1, 9}-8_{1, 8}  $ &  8 & 7  &   16482.250$\pm$0.002  &   1.43  & 1  \\
 $  9_{1, 9}-8_{1, 8}  $ & 10 & 9  &   16482.268$\pm$0.002  &   2.58  & 1  \\
 $  9_{0, 9}-8_{0, 8}  $ &  8 & 7  &   16699.940$\pm$0.002  &  -2.40  & 1  \\
 $  9_{0, 9}-8_{0, 8}  $ &  9 & 8  &   16699.952$\pm$0.002  &  -0.67  & 1  \\
 $  9_{0, 9}-8_{0, 8}  $ & 10 & 9  &   16699.962$\pm$0.002  &   1.45  & 1  \\
 $  9_{2, 8}-8_{2, 7}  $ &  9 & 8  &   16715.570$\pm$0.002  &   0.91  & 1  \\
 $  9_{2, 8}-8_{2, 7}  $ &  8 & 7  &   16715.609$\pm$0.002  &  -0.89  & 1  \\
 $  9_{2, 8}-8_{2, 7}  $ & 10 & 9  &   16715.627$\pm$0.002  &   6.96  & 1  \\
 $  9_{2, 7}-8_{2, 6}  $ &  9 & 8  &   16731.938$\pm$0.002  &  -0.20  & 1  \\
 $  9_{2, 7}-8_{2, 6}  $ &  8 & 7  &   16731.977$\pm$0.002  &  -1.17  & 1  \\
 $  9_{2, 7}-8_{2, 6}  $ & 10 & 9  &   16731.988$\pm$0.002  &  -0.36  & 1  \\
 $  9_{1, 8}-8_{1, 7}  $ &  9 & 8  &   16943.364$\pm$0.004  &   1.49  & 1  \\
 $  9_{1, 8}-8_{1, 7}  $ &  8 & 7  &   16943.364$\pm$0.004  &  -0.98  & 1  \\
 $  9_{1, 8}-8_{1, 7}  $ & 10 & 9  &   16943.382$\pm$0.002  &   1.53  & 1  \\
 $  2_{1, 2}-1_{0, 1}  $ &  1 & 0  &   18059.882$\pm$0.002  &  -0.45  & 1  \\
 $  2_{1, 2}-1_{0, 1}  $ &  2 & 2  &   18060.566$\pm$0.002  &  -0.32  & 1  \\
 $  2_{1, 2}-1_{0, 1}  $ &  2 & 1  &   18061.697$\pm$0.002  &  -0.81  & 1  \\
 $  2_{1, 2}-1_{0, 1}  $ &  1 & 1  &   18062.713$\pm$0.002  &   1.29  & 1  \\
 $ 10_{0,10}-9_{0, 9}  $ &  9 & 8  &   18551.073$\pm$0.002  &  -3.71  & 1  \\
 $ 10_{0,10}-9_{0, 9}  $ & 10 & 9  &   18551.085$\pm$0.002  &   0.45  & 1  \\
 $ 10_{0,10}-9_{0, 9}  $ & 11 &10  &   18551.093$\pm$0.002  &   1.71  & 1  \\
 $ 10_{2, 9}-9_{2, 8}  $ & 10 & 9  &   18572.031$\pm$0.002  &   0.86  & 1  \\
 $ 10_{2, 9}-9_{2, 8}  $ &  9 & 8  &   18572.058$\pm$0.002  &  -0.79  & 1  \\
 $ 10_{2, 9}-9_{2, 8}  $ & 11 &10  &   18572.070$\pm$0.002  &   1.83  & 1  \\
 $ 10_{2, 8}-9_{2, 7}  $ & 10 & 9  &   18594.518$\pm$0.002  &  -1.72  & 1  \\
 $ 10_{2, 8}-9_{2, 7}  $ &  9 & 8  &   18594.543$\pm$0.002  &  -4.45  & 1  \\
 $ 10_{2, 8}-9_{2, 7}  $ & 11 &10  &   18594.557$\pm$0.002  &   0.12  & 1  \\
 $ 10_{1, 9}-9_{1, 8}  $ & 10 & 9  &   18824.709$\pm$0.004  &   0.14  & 1  \\
 $ 10_{1, 9}-9_{1, 8}  $ &  9 & 8  &   18824.709$\pm$0.004  &  -0.98  & 1  \\
 $ 10_{1, 9}-9_{1, 8}  $ & 11 &10  &   18824.725$\pm$0.002  &   2.26  & 1  \\
 $  3_{1, 3}-2_{0, 2}  $ &  2 & 1  &   19841.616$\pm$0.002  &  -1.87  & 1  \\
 $  3_{1, 3}-2_{0, 2}  $ &  4 & 3  &   19841.943$\pm$0.002  &  -2.18  & 1  \\
 $  3_{1, 3}-2_{0, 2}  $ &  3 & 2  &   19842.165$\pm$0.002  &  -2.88  & 1  \\
 $ 11_{1,11}-0_{1,10}  $ & 11 &10  &   20142.342$\pm$0.002  &   1.49  & 1  \\
 $ 11_{1,11}-0_{1,10}  $ & 10 & 9  &   20142.342$\pm$0.002  &   1.68  & 1  \\
 $ 11_{1,11}-0_{1,10}  $ & 12 &11  &   20142.356$\pm$0.002  &   4.12  & 1  \\
 $ 11_{0,11}-0_{0,10}  $ & 10 & 9  &   20400.825$\pm$0.002  &  -2.97  & 1  \\
 $ 11_{0,11}-0_{0,10}  $ & 11 &10  &   20400.833$\pm$0.002  &  -1.06  & 1  \\
 $ 11_{0,11}-0_{0,10}  $ & 12 &11  &   20400.841$\pm$0.002  &   1.05  & 1  \\
 $ 11_{1,10}-0_{1, 9}  $ & 11 &10  &   20705.648$\pm$0.004  &  -2.66  & 1  \\
 $ 11_{1,10}-0_{1, 9}  $ & 10 & 9  &   20705.648$\pm$0.004  &  -3.02  & 1  \\
 $ 11_{1,10}-0_{1, 9}  $ & 12 &11  &   20705.662$\pm$0.002  &   0.31  & 1  \\
 $ 12_{0,12}-1_{0,11}  $ & 11 &10  &   22249.065$\pm$0.002  &  -2.53  & 1  \\
 $ 12_{0,12}-1_{0,11}  $ & 12 &11  &   22249.071$\pm$0.002  &  -1.30  & 1  \\
 $ 12_{0,12}-1_{0,11}  $ & 13 &12  &   22249.077$\pm$0.002  &  -0.54  & 1  \\
 $ 13_{0,13}-2_{0,12}  $ & 13 &12  &   24095.674$\pm$0.004  &  -0.89  & 1  \\
 $ 13_{0,13}-2_{0,12}  $ & 14 &13  &   24095.674$\pm$0.004  &  -5.64  & 1  \\
 $ 13_{0,13}-2_{0,12}  $ & 12 &11  &   24095.674$\pm$0.004  &   2.86  & 1  \\
 $ 14_{0,14}-3_{0,13}  $ & 14 &13  &   25940.525$\pm$0.004  &   1.84  & 1  \\
 $ 14_{0,14}-3_{0,13}  $ & 15 &14  &   25940.525$\pm$0.004  &  -2.52  & 1  \\
 $ 14_{0,14}-3_{0,13}  $ & 13 &12  &   25940.525$\pm$0.004  &   4.78  & 1  \\
 $ 15_{0,15}-4_{0,14}  $ & 14 &13  &   27783.508$\pm$0.004  &   4.79  & 1  \\
 $ 15_{0,15}-4_{0,14}  $ & 15 &14  &   27783.508$\pm$0.004  &   2.51  & 1  \\
 $ 15_{0,15}-4_{0,14}  $ & 16 &15  &   27783.508$\pm$0.004  &  -1.55  & 1  \\
 $ 17_{1,17}-6_{1,16}  $ &    &    &   31112.574$\pm$0.015  &   7.24  & 2  \\
 $ 17_{0,17}-6_{0,16}  $ &    &    &   31463.463$\pm$0.015  & -10.49  & 2  \\
 $ 17_{2,16}-6_{2,15}  $ &    &    &   31558.376$\pm$0.015  &  -6.50  & 2  \\
 $ 17_{3,15}-6_{3,14}  $ &    &    &   31593.484$\pm$0.020  & -47.78  & 2  \\
 $ 17_{3,14}-6_{3,13}  $ &    &    &   31596.159$\pm$0.016  &  -4.06  & 2  \\
 $ 17_{2,15}-6_{2,14}  $ &    &    &   31667.997$\pm$0.015  &   2.59  & 2  \\
 $ 17_{1,16}-6_{1,15}  $ &    &    &   31980.224$\pm$0.010  &  -9.75  & 2  \\
 $ 18_{1,18}-7_{1,17}  $ &    &    &   32939.203$\pm$0.010  &   6.65  & 2  \\
 $ 18_{0,18}-7_{0,17}  $ &    &    &   33300.280$\pm$0.010  &  -0.89  & 2  \\
 $ 18_{2,17}-7_{2,16}  $ &    &    &   33411.994$\pm$0.015  &  -3.94  & 2  \\
 $ 18_{3,16}-7_{3,15}  $ &    &    &   33453.015$\pm$0.017  &  -3.46  & 2  \\
 $ 18_{3,15}-7_{3,14}  $ &    &    &   33456.519$\pm$0.016  &  -4.13  & 2  \\
 $ 18_{2,16}-7_{2,15}  $ &    &    &   33541.670$\pm$0.015  &  24.42  & 2  \\
 $ 18_{1,17}-7_{1,16}  $ &    &    &   33857.014$\pm$0.010  & -10.27  & 2  \\
 $ 19_{1,19}-8_{1,18}  $ &    &    &   34765.294$\pm$0.015  &  24.28  & 2  \\
 $ 19_{0,19}-8_{0,18}  $ &    &    &   35134.847$\pm$0.020  & -29.61  & 2  \\
 $ 19_{2,18}-8_{2,17}  $ &    &    &   35265.143$\pm$0.020  &  -0.60  & 2  \\
 $ 19_{3,17}-8_{3,16}  $ &    &    &   35312.608$\pm$0.015  & -30.50  & 2  \\
 $ 19_{3,16}-8_{3,15}  $ &    &    &   35317.258$\pm$0.020  &  25.06  & 2  \\
 $ 19_{2,17}-8_{2,16}  $ &    &    &   35416.968$\pm$0.020  &   0.34  & 2  \\
 $ 19_{1,18}-8_{1,17}  $ &    &    &   35733.011$\pm$0.020  &  -3.80  & 2  \\
 $ 20_{1,20}-9_{1,19}  $ &    &    &   36590.744$\pm$0.010  & -24.55  & 2  \\
 $ 20_{0,20}-9_{0,19}  $ &    &    &   36967.205$\pm$0.015  &  -2.33  & 2  \\
 $ 20_{2,19}-9_{2,18}  $ &    &    &   37117.811$\pm$0.010  &  16.58  & 2  \\
 $ 20_{2,18}-9_{2,17}  $ &    &    &   37293.956$\pm$0.020  & -27.72  & 2  \\
 $ 21_{2,20}-0_{2,19}  $ &    &    &   38969.920$\pm$0.007  &  -5.53  & 2  \\
 $ 21_{2,19}-0_{2,18}  $ &    &    &   39172.697$\pm$0.008  &  -3.45  & 2  \\
 $ 22_{2,21}-1_{2,20}  $ &    &    &   40821.522$\pm$0.010  &   9.66  & 2  \\
 $ 22_{2,20}-1_{2,19}  $ &    &    &   41053.129$\pm$0.020  &  22.05  & 2  \\
 $ 22_{1,21}-1_{1,20}  $ &    &    &   41355.602$\pm$0.020  &  -2.82  & 2  \\
 $ 23_{1,23}-2_{1,22}  $ &    &    &   42063.663$\pm$0.005  &   0.85  & 2  \\
 $ 23_{0,23}-2_{0,22}  $ &    &    &   42450.325$\pm$0.005  &  -2.02  & 2  \\
 $ 23_{2,22}-2_{2,21}  $ &    &    &   42672.553$\pm$0.020  &  22.45  & 2  \\
 $ 23_{1,22}-2_{1,21}  $ &    &    &   43227.824$\pm$0.010  &  20.39  & 2  \\
\hline
\end{longtable}
\tablefoot{\\
        \tablefoottext{1}{Laboratory frequencies kindly provided by M. McCarthy.}\\
        \tablefoottext{2}{TMC-1 frequencies from Table \ref{obs_line_parameters}.}\\
}

\clearpage
\section{Column densities of CH$_3$C$_6$H and CH$_3$C$_5$N} \label{ch3c6h-5n}

\begin{figure*}[h]
\centering
\tiny
\includegraphics[width=1.0\textwidth]{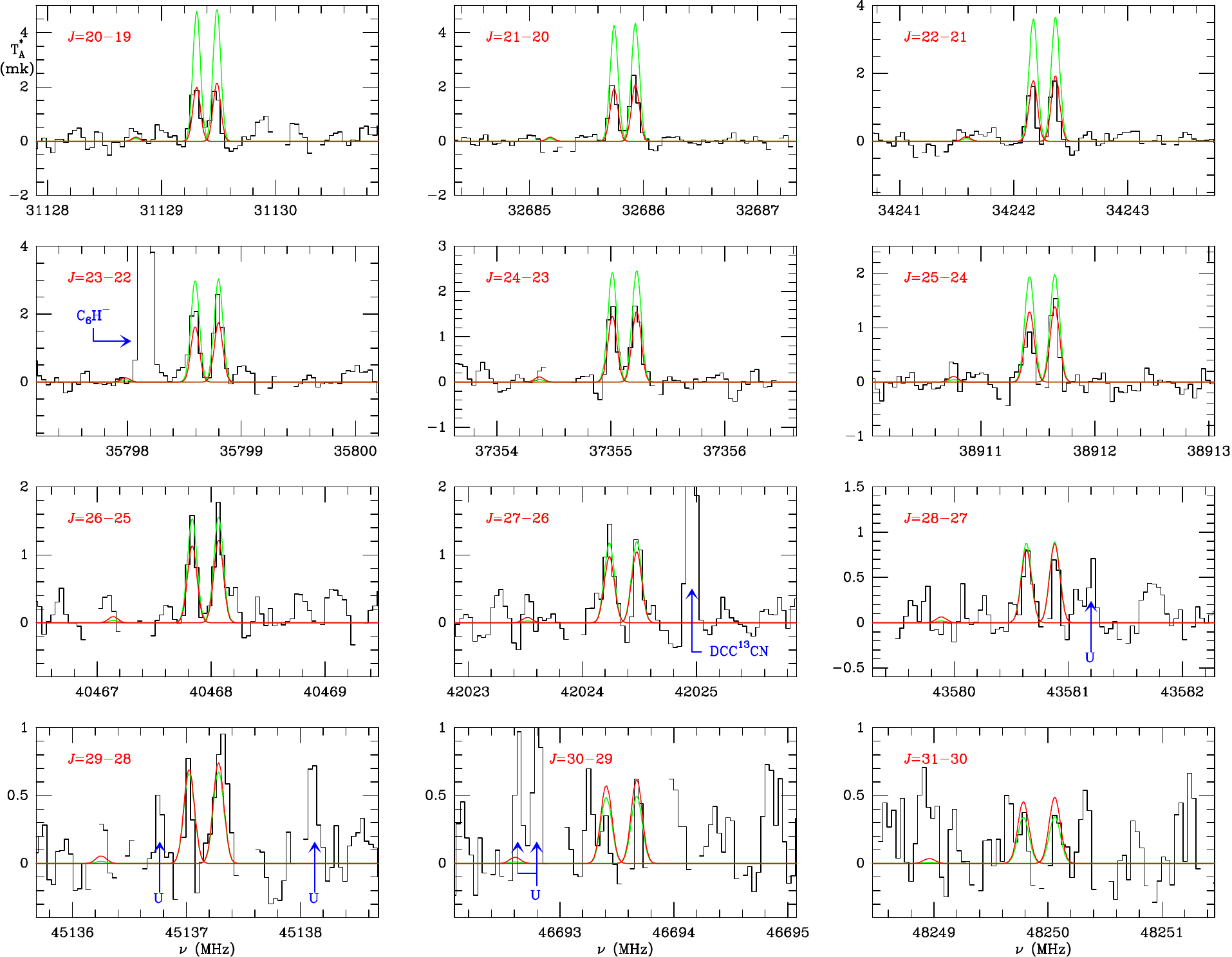}
\caption{Observed transitions of CH$_3$C$_6$H in TMC-1. The right component corresponds to K = 0 transitions and the left component to K = 1 transitions.
The abscissa corresponds to local standard of rest frequencies. Central frequencies and intensities for the observed lines
are given in Table \ref{obs_line_parameters}.
The ordinate is the antenna temperature, corrected for atmospheric and telescope losses, in milli Kelvin.
The quantum numbers of each transition are indicated
in the corresponding panel. The red lines show the computed synthetic spectrum for this species for $T_{\mathrm{rot}}$ = 9 K and
a column density of 7$\times$\once. The green line shows the computed synthetic spectrum 
assuming $T_{\mathrm{rot}}$ = 6 K and a $N$=3.1$\times$\doce \citep{Remijan2006}. Blanked channels correspond to negative features 
produced when folding the frequency-switched data.
}
\label{fig_ch3c6h}
\end{figure*}

\begin{figure*}
\centering
\includegraphics[width=1.0\textwidth]{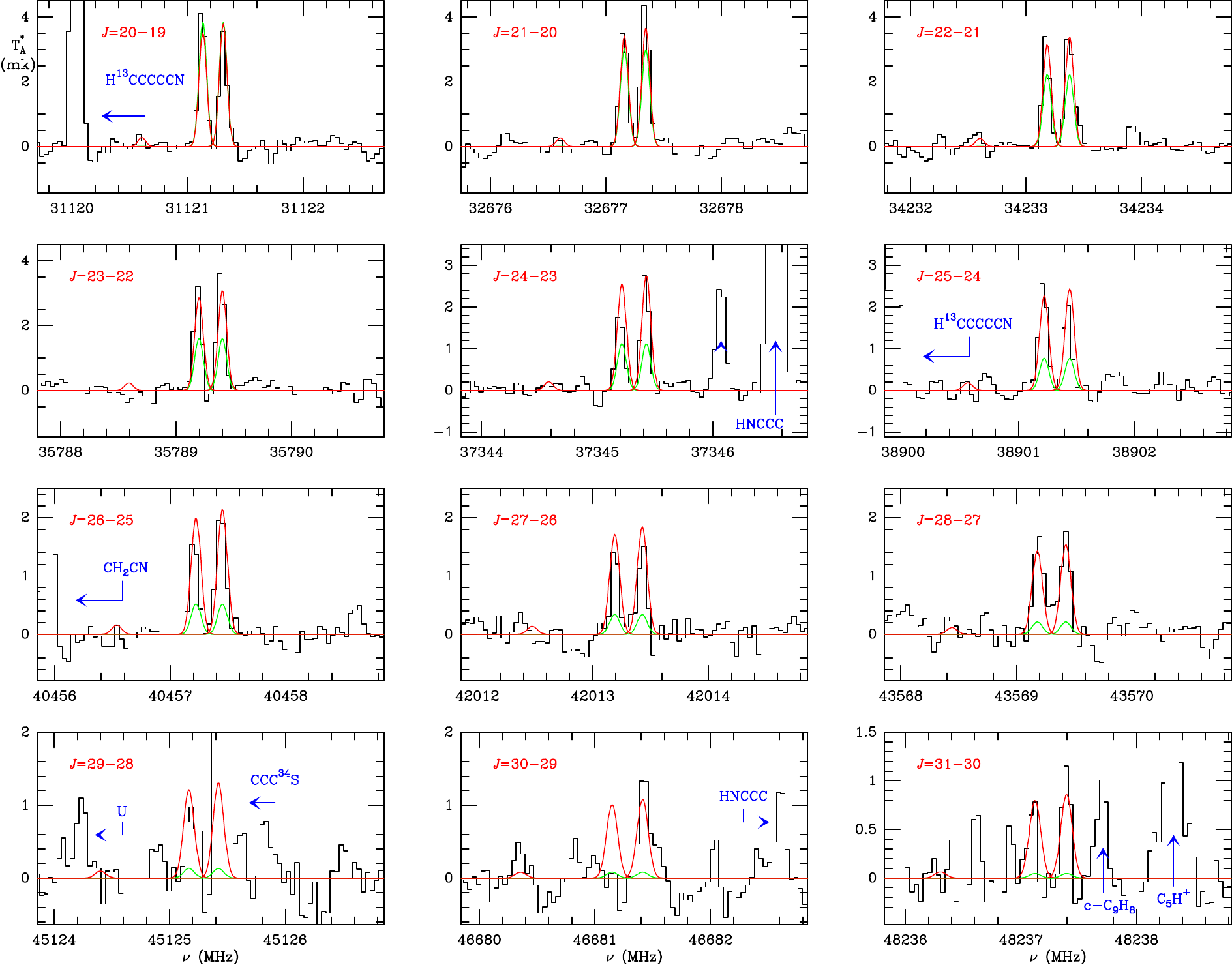}
\caption{Observed transitions of CH$_3$C$_5$N in TMC-1. The right component corresponds to K = 0 transitions and the left component to K = 1 transitions.
The abscissa corresponds to the rest frequency of the lines. Frequencies and intensities for the observed lines
are given in Table \ref{obs_line_parameters}.
The ordinate is the antenna temperature, corrected for atmospheric and telescope losses, in milli Kelvin.
The quantum numbers of each transition are indicated
in the corresponding panel. The red line shows the computed synthetic spectrum for this species for $T_{\mathrm{rot}}$ = 9 K and
a column density of 9.5$\times$\diez.  The green lines show the computed synthetic spectrum adopting
$T_{\mathrm{rot}}$ = 4 K and $N$=7.4$\times$\once \citep{Snyder2006}. Blanked channels correspond to negative features 
produced when folding the frequency-switched data.
}
\label{fig_ch3c5n}
\end{figure*}

In order to obtain abundance ratios between the different ethynyl and cyanide derivatives of CH$_2$CCH$_2$ in
a coherent and comprehensive way,
we derived the column densities of CH$_3$C$_6$H (an isomer of CH$_2$CCHC$_4$H)
and CH$_3$C$_5$N (an isomer of CH$_2$CCHC$_3$N) using the observed lines of these molecules 
within the QUIJOTE line survey. Both species have been previously detected towards \mbox{TMC-1} 
\citep{Remijan2006,Snyder2006}. These
authors reported a column density of  3.1\,$\times$\,10$^{12}$ cm$^{-2}$ and a rotational temperature of  6K 
for CH$_3$C$_6$H, and 
N = 7.4\,$\times$\,10$^{11}$ cm$^{-2}$ and $T_{\mathrm{rot}}$ = 4 K for CH$_3$C$_5$N.
In our data, we detected the rotational transitions $J$=20-19 up to $J$=31-30
for CH$_3$C$_6$H and CH$_3$C$_5$N ( having very similar rotational constants). The $K$=0,1 components of all
these lines (both species are symmetric rotors) are detected except the $K$=1 component of the $J$=30-29 
transition of CH$_3$C$_5$N, which we
explain as possibly due to an overlap with a negative feature produced in the folding of the frequency switching
data. The observed transitions are shown on Fig. \ref{fig_ch3c6h} for CH$_3$C$_6$H and Fig. \ref{fig_ch3c5n} for
CH$_3$C$_5$N.
No $K$=2 lines are detected, which is reasonable taking into account the kinetic temperature of the
cloud (10 K) and the high energy of the $K$=2 levels (from 38.9 for $J_u$=20 up to 57.3 K for  $J_u$=30).

We derived the column densities and the rotational temperatures
of CH$_3$C$_6$H and CH$_3$C$_5$N using a model line fitting procedure (see 
Section \ref{results} in this work as well as \citealt{Cernicharo2021d}). We adopt a source of uniform brightness 
with a diameter of 80$''$ \citep{Fosse2001,Cernicharo2022} and a linewidth of 0.6 \kms. The best model for our CH$_3$C$_6$H lines provides 
$N$=(7$\pm$0.7)\,$\times$\,10$^{11}$ cm$^{-2}$ and $T_{\mathrm{rot}}$ =9$\pm$0.5 K. 
The computed synthetic line profiles using these values are shown on Fig. \ref{fig_ch3c6h} as red lines.
The derived rotational temperature and column density are rather different 
from those derived previously by \citet{Remijan2006}. This difference is certainly 
due to the larger energy range covered by our data, which renders the line intensities very
dependent on the rotational temperature. In Fig. \ref{fig_ch3c6h}, we show in green the expected line profiles
if we adopt the rotational temperature and column density derived by \citet{Remijan2006}. The low-$J$ transitions
overestimate the observed intensities, while a reasonable agreement is found for $J_u\ge$27. Using our derived parameters, we reproduce the intensities of the low-$J$ transitions observed by \citet{Remijan2006} within a factor of 1.5.

For CH$_3$C$_5$N, we derived $N$=(9.5$\pm$0.9)\,$\times$\,10$^{10}$ cm$^{-2}$ and 
$T_{\mathrm{rot}}$ = 9$\pm$0.5 K. The predicted synthetic lines for this species are shown on red in Fig \ref{fig_ch3c5n}. 
These values are also different from those derived by \citet{Snyder2006}, $T_{\mathrm{rot}}$ = 4 K and $N$=7.4$\times$\once.
With such a low value for the rotational temperature, it is impossible to reproduce the observed intensities
of our lines. We show in green in Fig. \ref{fig_ch3c5n} the computed synthetic spectrum of CH$_3$C$_5$N
adopting the parameters of \citet{Snyder2006}. A good match is obtained for the transitions
$J$=20-19 and $J$=21-20 but a clear underestimation of the intensity is observed for lines with
higher $J$. Using our $T_{\mathrm{rot}}$ and column density for this species, we reproduce  the intensities observed for the low-$J$ transitions by \citet{Snyder2006} within a factor of 2. 

Adopting a H$_2$ column density for TMC-1 of 10$^{22}$ cm$^{-2}$ \citep{Cernicharo1987}, we obtain an abundance 
of (7$\pm$0.7)\,$\times$\,10$^{-11}$ for CH$_3$C$_6$H, and  
of (9.5$\pm$0.9)\,$\times$\,10$^{-12}$ for CH$_3$C$_5$N. 

The relative abundance of the isomers CH$_2$CCHC$_4$H and CH$_3$C$_6$H is, hence, 3.1$\pm$0.3. For the pair of isomers
CH$_2$CCHC$_3$N and CH$_3$C$_5$N, their abundance ratio is 1.3$\pm$0.2.

\twocolumn
\end{appendix}

\end{document}